\documentclass[journal]{IEEEtran}

\usepackage{cite}
\usepackage{easybmat}
\usepackage{stfloats}
\usepackage{amsmath,amssymb,amsfonts}
\usepackage{graphicx}
\usepackage{textcomp}
\usepackage[linesnumbered,lined,boxed,commentsnumbered, ruled]{algorithm2e}
\usepackage{epstopdf}
\usepackage{balance}
\usepackage{caption}
\usepackage[printonlyused,withpage]{acronym}
\usepackage{subcaption}
\usepackage{algorithmic}
\usepackage{xcolor}
\usepackage[letterpaper, left=0.7in, right=0.7in, bottom=1in, top=0.7in]{geometry}

\newtheorem{prop}{Proposition}

\newtheorem{rmk}{Remark}

\DeclareFontFamily{U}{mathx}{}
\DeclareFontShape{U}{mathx}{m}{n}{ <-> mathx5 }{}
\DeclareSymbolFont{mathx}{U}{mathx}{m}{n}
\DeclareFontSubstitution{U}{mathx}{m}{n}

\DeclareMathAccent{\widecheck}{0}{mathx}{"71}

\newenvironment{proof}{ \paragraph*{Proof}}{\hfill$\square$}

\newcommand{\mrm}[1]{\ensuremath{\text{#1}}}
\renewcommand{\vec}{\ensuremath{\mrm{vec}}}
\newcommand{\diag}{\ensuremath{\mrm{diag}}}
\newcommand{\bs}[1]{\ensuremath{\boldsymbol{#1}}}
\newcommand{\comment}[1]{}

\renewcommand{\baselinestretch}{.994} 

\acrodef{ADC}{analog-to-digital converter}

\acrodef{IID}{independent and identically distributed}
\acrodef{MIMO}{multiple-input multiple-output}
\acrodef{MU-MIMO}{multi-user \ac{MIMO}}
\acrodef{UE}{user equipment}
\acrodef{BS}{base station}
\acrodef{SNR}{signal-to-noise ratio}
\acrodef{SINR}{signal-to-interference-plus-noise ratio}
\acrodef{SVD}{singular-value decomposition}
\acrodef{EVD}{eigenvalue decomposition}
\acrodef{CSI}{channel state information}
\acrodef{RIS}{reconfigurable intelligent surface}
\acrodef{BD-RIS}{beyond-diagonal reconfigurable intelligent surface}
\acrodef{BBU}{baseband unit}
\acrodef{FRIS}{fully-reconfigurable intelligent surface}
\acrodef{DEP}{\textit{dimensionally-efficient parametrization}}
\acrodef{AWGN}{additive white Gaussian noise}
\acrodef{MSE}{mean squared error}
\acrodef{ICA}{independent component analysis}
\acrodef{RNN}{recurrent neural networks}
\begin{document}

\title{Dimensionally-Efficient Transmission and Storage of Unitary Matrices}
\author{Juan Vidal Alegr\'{i}a\\
\thanks{This work was supported by the Swedish Foundation of Strategic Research (CHI19-0001).\par J. Vidal Alegr\'{i}a is with Lund University, SE-22100 Lund, Sweden (corresponding email: juan.vidal\_alegria@eit.lth.se).} }

\maketitle
\begin{abstract} 
Unitary matrices are the basis of a large number of signal processing applications. In many of these applications, finding ways to efficiently store, and even transmit these matrices, can significantly reduce memory and throughput requirements. In this work, we study the problem of efficient transmission and storage of unitary matrices. Specifically, we explicitly derive a dimensionally-efficient parametrization (DEP) for unitary matrices that allows identifying them with sequences of real numbers, where the dimension coincides with the dimension of the unitary group where they lie. We also characterize its inverse map that allows retrieving the original unitary matrices from their DEP. The proposed approach effectively allows halving the dimension with respect to naively considering all the entries of each unitary matrix, thus reducing the resources required to store and transmit these matrices. Furthermore, we show that the sequence of real numbers associated to the proposed DEP is bounded, and we delimit the interval where these numbers are contained, facilitating the implementation of quantization approaches with limited distortion. On the other hand, we outline ways to further reduce the dimension of the DEP when considering more restrictive constraints for matrices that show up in certain applications. The numerical results showcase the potential of the proposed approach in general settings, as well as in three specific applications of current interest for wireless communications research.

\end{abstract}
\begin{IEEEkeywords}
Unitary matrices, dimensionally-efficient parametrization (DEP), multiple-input multiple-output (MIMO), reconfigurable surfaces, distributed-MIMO (D-MIMO).
\end{IEEEkeywords}

\section{Introduction}\label{section:intro}
\IEEEPARstart{U}{nitary} matrices have a crucial role in signal processing, finding application in a long list of diverse fields that include wireless communications \cite{molisch}, biomedical engineering \cite{akay_biomed}, microwave theory \cite{pozar}, and quantum computation \cite{nielsen}. Unitary transformations have unique properties which allow them to preserve norms and inner products, while they can be inverted at low complexity by simple conjugate transposition. These and other features make them one of the basic building blocks for spectral analysis, quantum processing, \ac{ICA}, and other fundamental signal processing techniques.

Many of the applications that employ unitary matrices require storing these matrices so that they can be accessible at different instances. In the context of quantum computing, finding ways to physically store unitary matrices in quantum systems is a research area in itself \cite{quant_unit_strg,quant_unit_strg2}. However, the current work focuses on the storage of unitary matrices using classical computing systems, specifically in the context of the widely employed digital technology.\footnote{Classical computing is still the main technology used to perform signal processing tasks, while there seems to be a necessity for their coexistence with quantum computing systems towards future technological advances \cite{nielsen}.} In this sense, the goal is to store unitary matrices effectively while using the least possible memory resources, e.g., mapping unitary matrices with the shortest binary sequences that allow recovering them under given error. To the best of our knowledge, this problem has not been explicitly studied in previous literature.

The reduced dimensionality of the unitary group allows parameterizing unitary matrices through a sequence of real numbers half as long as the one required to parameterize arbitrary complex matrices. Hence, a straightforward way to store unitary matrices efficiently is to define a suitable \ac{DEP}, i.e., a parametrization having minimum number of dimensions, and to store instead the resulting sequence of minimum dimension that allows retrieving each unitary matrix without loss. In \cite{gen_gell_mann} a \ac{DEP} is defined to represent qudits as a sequence of $N^2-1$ real numbers, which coincides to the real dimension of the special unitary group (a restricted version of the unitary group) where the qudits lie. In \cite{learning_un}, a similar \ac{DEP} is considered for towards the goal of training \acp{RNN} with unitary transition matrices. In this work, we show how to practically exploit such parametrizations to efficiently store unitary matrices.

In some applications, specially when dealing with wireless communication systems, it may even be necessary to send unitary matrices to other remote locations so that different systems can employ them to perform their tasks. For example, in order to achieve full channel capacity in \ac{MIMO} systems, the transmitter has to precode its symbols using the right unitary matrix from the \ac{SVD} of the \ac{MIMO} channel matrix. This requires to consider \ac{CSI} feedback strategies to transmit the respective unitary matrix, which is typically only known at the receiver where the channel is estimated. This has been a topic widely studied in the literature with numerous solutions which may constitute the most relevant state-of-the-art for our study \cite{givens_CSI, givens2_spatial_mat,stiefel_pred_quant,diff2}. In this context, \cite{givens_CSI} considers a \ac{DEP} based on Givens rotations for unitary matrices up to a diagonal phase uncertainty. This \ac{DEP} is then used to reduce the \ac{CSI} feedback rate by performing optimized quantization upon it. A similar approach based on Givens rotations is considered in \cite{givens2_spatial_mat}, which further proposes differential schemes to update the respective \ac{CSI}. Other differential techniques for transmission of unitary matrices have been considered in \cite{stiefel_pred_quant,diff2}.

Due to the intrinsic equivalence between the rate required to transmit a unitary matrix, and the memory required to store it \cite{inf_th}, many of the methods for transmitting unitary matrices can be directly reused for the storage counterpart. However, the main limitation of the approaches in \cite{givens_CSI, givens2_spatial_mat,stiefel_pred_quant,diff2} lie in the fact that they are tailored to the \ac{CSI} feedback application, e.g., by assuming specific channel models with given spatial distributions or time-correlation structures, or by ignoring the diagonal phase uncertainty of the transmitted unitary matrices. Moreover, the differential techniques \cite{givens2_spatial_mat,stiefel_pred_quant,diff2} have the extra limitation that they still require an initial full-estimate of the respective unitary matrix. Thus, finding techniques to store and transmit unitary matrices in general settings remains an open problem. 

In the context of modern wireless communications research, several promising enabling technologies have been proposed towards future generations of wireless systems. Some of these enabling technologies depend strongly on the possibility to effectively store and transmit unitary matrices. For example, the recently proposed \ac{BD-RIS} \cite{bd-ris,bd-ris1,nr-BD-RIS}, which extends the widespread 
\ac{RIS} \cite{en_eff_RIS,RIS} by including inter-element reconfigurable connections to further enhance performance, leads to reflection matrices with unitary constraints. Since the optimal reflection matrices are typically computed at a transceiver node with channel estimation capabilities, forwarding the computed matrices to the \ac{BD-RIS}, where they are physically applied, may directly benefit from efficient methods to send and store unitary matrices. Another recent proposal is to use lossless microwave analog computing networks for \ac{MIMO} processing \cite{anal_comp,MIMO_anal_comp,asilomar24}, which are characterized by scattering matrices with unitary constraints. In this case, sharing these matrices between the nodes where they are computed and the nodes where they are applied may also benefit from efficient storage and transmission of unitary matrices. To the best of our knowledge, this issue has been largely overlooked in existing literature.

The main contributions of this paper are summarized next:
\begin{itemize}
    \item We derive through explicit transformations a \ac{DEP} for unitary matrices to map arbitrary unitary matrices with sequences of real numbers with minimum dimension.
    \item We study the time-complexity associated to computing the proposed \ac{DEP}, as well as to retrieving the original unitary matrix from it, and conclude that the order does not increase compared with performing \ac{SVD}---which is often a pre-requisite for the considered applications.
    \item We show that the proposed \ac{DEP} maintains the property of boundedness, and derive a range within which the resulting real numbers should be contained.
    \item We outline how to adapt the proposed \ac{DEP} to incorporate extra constraints such as symmetry, real-valuedness (i.e., for orthogonal matrices), or phase uncertainty.
    \item We study three wireless communications applications, specifically in the context of modern \ac{MIMO} technologies as mentioned above, and we show how they may benefit from the considered approach.
    \item We numerically analyze the potential of the proposed \ac{DEP} in the task of storage and transmission of unitary matrices in general tasks, as well as in the specific wireless communication applications previously mentioned.
\end{itemize}

Some of the methods hereby exploited may appear trivial for readers familiar with Lie theory \cite{lie_groups}. However, an important contribution of this work is to facilitate their practical exploitation, and to point out their significance in the context of storage and transmission of unitary matrices---specially in the context of wireless communications. On the other hand, the proof of boundedness of the proposed \ac{DEP} for unitary matrices poses, to the best of our knowledge, a novel theoretical contribution, further providing a useful explicit range for the values of the entries of this \ac{DEP}. The previous result may be exploited to implement simple quantization methods with bounded error. 




\section{Problem formulation}
Consider the transmission/storage\footnote{In this work we can interchange the word transmission (transmit) with storage (store), since both information rate and level of compression are similarly influenced by the parametrization of the input \cite{inf_th}.} of an $N\times N$ unitary matrix, $\bs{U}\in \mathcal{U}(N)$, where $\mathcal{U}(N)$ denotes the unitary group. The straightforward approach to parameterize $\bs{U}$ is through the $N^2$ complex numbers composing it. Since each complex number may be characterized by two real numbers (its real and imaginary part), this would correspond to characterizing $\bs{U}$ by $2N^2$ real numbers. Thus, in order to transmit $\bs{U}$ we could simply send the sequence of $2N^2$ real numbers composing it, e.g., through $2N^2$ uses of a scalar channel. This naive approach would still be the most dimensionally-efficient way to transmit an arbitrary complex matrix $\bs{A}\in \mathbb{C}^{N\times N}$, where we assume no specific structure on $\bs{A}$. The reason is that the space $\mathbb{C}^{N\times N}$ of $N\times N$ complex matrices is composed of $2N^2$ real dimensions, so in order to differentiate $\bs{A}$ from any other arbitrary point in $\mathbb{C}^{N\times N}$ we would need to parameterize it with at least $2N^2$ real numbers. However, it is well known that $\mathcal{U}(N)$ corresponds to a topological (Riemannian)\footnote{In order to consider $\mathcal{U}(N)$ as a Riemannian manifold it should further be equipped with a Riemannian metric. However, this work does not directly require such tool.} manifold of $N^2$ real dimensions \cite{riemann}. Hence, it should be possible to parameterize without loss any unitary matrix as a sequence of $N^2$ real numbers, hereby referred to as a \ac{DEP} for $\mathcal{U}(N)$.

The notion of $\mathcal{U}(N)$ being considered a topological (Riemannian) manifold of (real) dimension $N^2$ inherently means that there exist local homeomorphic (bijective) maps between $\mathcal{U}(N)$ and $\mathbb{R}^{N^2}$ \cite{riemann}. However, even if these maps (also termed local charts/coordinates) can be explicitly defined, the term "local" means that each of these may only apply to neighborhoods around specific elements of $\mathcal{U}(N)$ (and their respective image within $\mathbb{R}^{N^2}$). Due to the nature of the problem under consideration, it is in fact desirable to have a more general construction: a unique map that can be applied to any element of $\mathcal{U}(N)$ and leads to distinguishable elements within $\mathbb{R}^{N^2}$. This way, if we have one transmitter and one receiver (conversely one system storing a unitary matrix and another one accessing it) they should both have a common function to translate from unitary matrices to their \acp{DEP} without having to know further details about the neighborhoods around which these translations would work. An initial goal of this work is thus to explicitly define a \ac{DEP} for $\mathcal{U}(N)$, which consists of an injective function 
\begin{equation}\label{eq:fun_UR}
    \bs{f}: \mathcal{U}(N)\to \mathbb{R}^{N^2},
\end{equation}
which allows to uniequivocally identify an arbitrary unitary matrix $\bs{U}\in \mathcal{U}(N)$ with a real vector $\bs{v}=\bs{f}(\bs{U})\in \mathbb{R}^{N^2}$.\footnote{Note that surjectivity can always be enforced by restricting the output domain since we may assume that we will always start with a given unitary matrix, leading to a well-defined inverse function $\bs{f}^{-1}:\bs{f}(\mathcal{U}(N))\to \mathcal{U}(N)$.}

Another valuable property of the unitary group is its compactness \cite{lie_groups,riemann}. This implies that $\mathcal{U}(N)$ is a closed and bounded (under the standard metric) subspace of $\mathbb{C}^{N\times N}$, i.e., in the same way as the interval $[0,1]$ is a closed and bounded subspace of $\mathbb{R}$. The reason why this property is important is that, if we consider digital systems, in order to transmit/store a real number $r\in \mathbb{R}$ we first need to be able to translate it to a finite binary sequence. This is usually done by quantization \cite{quant_surv}, which corresponds to approximating $r$ with the closest value $\hat{r}\in \mathcal{Q}$ within a finite subset $\mathcal{Q}\subset \mathbb{R}$ ($\vert \mathcal{Q} \vert =M$). From the finite property, the quantized subset for any quantization process is inevitably contained within a bounded (compact) interval, i.e., $\mathcal{Q}\subset[a,b]$. Thus, if the real number $r$ takes values much larger than $b$ (or much smaller than $a$), the quantization process would incur significant distortion and information loss due to clipping effects \cite{proakis}. On the other hand, if we know that $r$ is contained within a bounded subset of $\mathbb{R}$, e.g., $r\in[a',b']$, we can avoid clipping altogether by adjusting accordingly the quantization interval. Hence, it is desirable that our \ac{DEP} $\bs{f}(\bs{U})\in \mathbb{R}^{N^2}$ maintains the boundedness property inherent to the unitary group, i.e., such that $\bs{f}(\bs{U})\in [\tau_{\min},\tau_{\max}]^{N^2}$. This would allow us to limit the distortion associated to quantizing $\bs{f}(\bs{U})$ since we could adjust the quantization intervals to avoid/limit clipping effects. Moreover, we would like to characterize the tightest interval $[\tau_{\min},\tau_{\max}]$ such that the image of $\mathcal{U}(N)$ under $\bs{f}$ approximately coincides with $[\tau_{\min},\tau_{\max}]^{N^2}$. This would allow us to efficiently quantize the entries of $\bs{f}(\bs{U})\in [\tau_{\min},\tau_{\max}]^{N^2}$ by tightly adjusting the quantization intervals.

Altogether, this work aims at defining an injective function
\begin{equation}\label{eq:fun_comp}
    \bs{f}: \mathcal{U}(N)\to [a,b]^{N^2},
\end{equation}
with a codomain $[a,b]^{N^2}$ as tight as possible to its image $\bs{f}(\mathcal{U}(N))$. Ideally, we would like $\bs{f}$ to be also surjective (or to be able to precisely delimit its image) so that its codomain can be made equal to its image, allowing for perfect adjustment of the dynamic range of the quantizers potentially employed to digitalize the outputs of $\bs{f}$. However, ensuring surjectivity is a complex task which only offers a limited gain in terms of quantization resolution. Thus, we consider that surjetivity is not a formal requirement.

\section{Unitary Matrix Dimensionally-Efficient Parametrization}\label{section:unit}
\subsection{The exponential map for $\mathcal{U}(N)$}
\label{ss:exp_m}
Let us start by introducing the exponential map. For general Lie groups, the exponential map is defined as a map between the Lie algebra associated to the Lie group and the group itself. Alternatively, we can say that the exponential map links the tangent plane at the identity element with the Lie group, which gives a more geometric interpretation (one may think of a sphere being unfolded into its tangent plane). In the case of the unitary group, we can write
\begin{equation}\label{eq:exp_map}
    \exp: \mathfrak{u}(N) \to \mathcal{U}(N),
\end{equation}
where $\mathfrak{u}(N)$ corresponds to the unitary Lie group, which is defined as \cite{riemann,lie_groups,traian}.
\begin{equation}\label{eq:unit_Lalg}
    \mathfrak{u}(N) \triangleq T_{\mathbf{I}_N}\mathcal{U}(N)=\{\boldsymbol{X}\in \mathbb{C}^{N\times N}:\bs{X}^\mrm{H}+\bs{X}=\mathbf{0} \},
\end{equation}
with $T_{\mathbf{I}_N}\mathcal{U}(N)$ denoting the tangent plane to $\mathcal{U}(N)$ at the identity element, $\mathbf{I}_{N}$. From \eqref{eq:unit_Lalg}, we can note the correspondence between $\mathfrak{u}(N)$ and the space of skew-Hermitian matrices \cite{riemann,lie_groups}.

The exponential map is a useful tool to link a Lie algebra with a Lie group. However, the premise of this work is that, starting with a unitary matrix, we would like to link it to a \ac{DEP}. To this end, we should also consider the inverse exponential map, which allows mapping a Lie group to its Lie algebra. For the unitary group, we can write
\begin{equation}\label{eq:log_map}
    \exp^{-1}: \mathcal{U}(N) \to \mathfrak{u}(N).
\end{equation}

A Lie algebra corresponds to a real vector space of the same dimension as its Lie group, endorsing it with nice properties to operate with its elements by exploiting standard notions from linear algebra \cite{riemann,lie_groups}. On the other hand, the exponential map for any connected, compact matrix Lie group (as the unitary group) is always surjective. This means that the inverse exponential map, defined in \eqref{eq:log_map} for the unitary group, is injective. Thus, this map fulfills our initial requirement for unequivocally identifying unitary matrices with elements of their Lie algebra. We next focus on providing explicit expressions for computing the exponential map, and the inverse exponential map, in the context of $\mathcal{U}(N)$.

For any matrix Lie group (as $\mathcal{U}(N)$), the exponential map coincides with the matrix exponential \cite{riemann,lie_groups}. Conversely, we may identify the inverse exponential map with the matrix logarithm (we can substitute $\exp^{-1}$ in \eqref{eq:log_map} with $\log$). Note that, within the convergence region of the matrix logarithm, we have that $\exp(\log(\bs{A}))=\bs{A}$ \cite[Theorem~2.8]{lie_groups}. Since both unitary matrices and skew-Hermitian matrices fall within the class of normal matrices \cite{mat_anal} (and are thus unitarily diagonalizable), we can give an explicit expression for the matrix exponential of a skew-Hermitian matrix, and for the matrix logarithm of a unitary matrix, based on their \acp{EVD}. Specifically, for $\bs{X}\in \mathfrak{u}(N)$ we have
\begin{equation}\label{eq:exp_skh}
    \exp(\bs{X}) = \bs{U}_{\bs{X}}\exp(\bs{\Lambda}_{\bs{X}})\bs{U}_{\bs{X}}^\mrm{H},
\end{equation}
and for $\bs{U}\in \mathcal{U}(N)$ we have
\begin{equation}\label{eq:log_unit}
    \log(\bs{U}) = \bs{U}_{\bs{U}}\log(\bs{\Lambda}_{\bs{U}})\bs{U}_{\bs{U}}^\mrm{H},
\end{equation}
where we have considered the \acp{EVD} $\bs{X} = \bs{U}_{\bs{X}}\bs{\Lambda}_{\bs{X}}\bs{U}_{\bs{X}}^\mrm{H}$ and $\bs{U} = \bs{U}_{\bs{U}}\bs{\Lambda}_{\bs{U}}\bs{U}_{\bs{U}}^\mrm{H}$, with $\bs{\Lambda}_{\bs{X}}$ and $\bs{\Lambda}_{\bs{U}}$ corresponding to diagonal matrices, and $\bs{U}_{\bs{X}}$ and $\bs{U}_{\bs{U}}$ corresponding to unitary matrices. Note that, for diagonal matrices, the exponential/logarithm is defined as the traditional complex exponential/logarithm applied to the diagonal entries. On the other hand, for unitary matrices that do not fulfill the explicit convergence criterion of the matrix logarithm, given by $\Vert\bs{U}-\mathbf{I}_N\Vert<1$ \cite[Theorem~2.8]{lie_groups}, we can still have a well defined inverse exponential map by considering the complex logarithm \cite{complex_anal} in \eqref{eq:log_unit}, which allows having real negative diagonal entries in $\bs{\Lambda}_{\bs{U}}$ while ensuring $\exp(\log(\bs{U}))=\bs{U}$. The only remaining cases leading to an ill-defined logarithm in \eqref{eq:log_unit} would then correspond to when $\bs{\Lambda}_{\bs{U}}$ has diagonal entries equal to 0. However, this can never happen when considering unitary matrices since their eigenvalues always lie on the complex unit circle \cite[Section 4.5]{unit_prop},\cite{eig_unit_herm}, i.e., the diagonal elements of $\bs{\Lambda}_{\bs{U}}$ are always unimodular (thus non-zero).
\begin{rmk}
    The relation between unitary matrices and Skew-Hermitian matrices may be understood from the standard scalar/element-wise complex exponential and logarithm. Unitary matrices have eigenvalues which are modulus-1 complex numbers, while Skew-Hermitian matrices (which may be seen as Hermitian matrices scaled by the imaginary unit) have eigenvalues which are purely imaginary \cite{eig_unit_herm}. Hence, \eqref{eq:log_unit} and \eqref{eq:eig_skew} map the elements from $\mathfrak{u}(N)$ with the elements from $\mathcal{U}(N)$ through their eigenvalues, in the same way that the standard complex logarithm and exponential map modulus-1 complex numbers with purely imaginary numbers.
\end{rmk}

\begin{rmk}
    An alternative injective map between Lie algebra $\mathfrak{u}(N)$ and the Lie Group $\mathcal{U}(N)$ could be the Cayley transform, as considered in \cite{cayley_codes} for the alternative problem of generating unitary space-time codes. However, this transform would not work for cases where the input unitary matrix has any eigenvalues at -1, which limits its applicability. Moreover, the resulting image would not be bounded since when inputting a unitary matrix with eigenvalues close to -1 the resulting Skew-Hermitian matrix would have the respective eigenvalues approaching infinity, reducing practicality since having a compact image in \eqref{eq:fun_comp} would not be possible.
\end{rmk}

We have now defined the maps that allow unequivocally translating unitary matrices to elements of the real vector space $\mathfrak{u}(N)$, and vice-versa. Our next goal is to exploit the real vector space property of $\mathfrak{u}(N)$ to parameterize its elements as dimensionally-efficient sequences of real numbers within a well-defined basis. 

\subsection{A basis for $\mathfrak{u}(N)$}
\label{ss:basis}
Let us consider the standard Hilbert-Schmidt inner product $\langle\bs{A},
\bs{B}\rangle=\mrm{trace}(\bs{A}^\mrm{H}\bs{B})$, which induces the squared Frobenius norm for $\bs{A}=\bs{B}$.\footnote{The Frobenius norm is specially interesting here since it ensures coherence between the norm of a matrix and the norm of its vectorized form, i.e., $\Vert\bs{A}\Vert_\mrm{F}=\Vert\vec(\bs{A})\Vert$.} In this section, we will propose an orthonormal basis for $\mathfrak{u}(N)$ under the previous considerations. This basis will allow us to extract the $N^2$ real coordinates associated to each element $\bs{X}\in \mathfrak{u}(N)$, by simply applying a unitary transformation on the vectorized $\vec (\bs{X})$.

As noted in Section~\ref{ss:exp_m}, $\mathfrak{u}(N)$ corresponds to a real vector space which may be identified with the space of skew-Hermitian matrices \eqref{eq:unit_Lalg}. The generalized Gell-Mann matrices \cite{gen_gell_mann,gen_gell_mann2}, offer an orthogonal basis for the space of $N\times N$ Hermitian traceless matrices. By further including the identity matrix $\mathbf{I}_N$ as an extra basis element, the complete basis would actually span the whole space of $N\times N$ Hermitian matrices (traceless or not)\cite{gen_gell_mann2}. On the other hand, any skew-Hermitian matrix can be generated as a Hermitian matrix scaled by the complex unit $\mrm{j}$, since this exchanges the properties of the real and imaginary parts. Note that the real part of a Hermitian matrix is symmetric and its imaginary part is antisymmetric, while the converse is true for skew-Hermitian matrices. Hence, we can obtain an orthogonal basis for $\mathfrak{u}(N)$ by including the generalized Gell-Mann matrices, together with the identity matrix, each scaled by $\mrm{j}$. We can further divide each basis element by its Frobenius norm to get a normalized basis within the considered framework. This gives the basis
\begin{equation}
    \mathcal{B}=\{\bs{B}_1,\dots,\bs{B}_{N^2}\},
\end{equation}
whose elements are defined in \eqref{eq:basis}. The last three lines of \eqref{eq:basis} correspond to the scaled generalized Gell-Mann matrices consisting of $N-1$ imaginary diagonal matrices, $N(N-1)/2$ imaginary symmetric matrices, and $N(N-1)/2$ real anti-symmetric matrices. Note that the arrangement of the basis elements is irrelevant as long as the basis contains all the $N^2$ mutually-orthogonal elements. However, a convention should be adopted so that different systems, e.g., a transmitter and a receiver, may coherently retrieve unitary matrices from their \acp{DEP}. For the symmetric and anti-symmetric elements in the last two rows of \eqref{eq:basis}, one possible arrangement is obtained by row traversing the upper triangular part, which gives
\begin{subequations}\label{eq:k_n_l_n}
\begin{equation}\label{eq:k_n}
    k(n) = \left\lfloor \frac{2N-1-\sqrt{(2N-1)^2-8n}}{2} \right\rfloor+1,
\end{equation}
\begin{equation}
    l(n) = k(n)-\big(n \bmod (N-k_n)\big).
\end{equation}
\end{subequations}
Note that $k(n)$ in \eqref{eq:k_n} simply corresponds to a sequence of $N-1$ $1$s, followed by $N-2$ $2$s, and so forth until reaching the last element equaling $N-1$.

\begin{figure*}
\begin{equation}\label{eq:basis}
    \bs{B}_n = \left\{\begin{matrix}
    \frac{\mrm{j}}{\sqrt{N}} \mathbf{I}_N,& n=1\\
    \frac{\mrm{j}}{\sqrt{n(n-1)}}\left(\sum_{i=1}^{n-1}\bs{E}_{i,i}-n\bs{E}_{n,n}\right),& n=2,\dots,N\\
    \frac{\mrm{j}}{\sqrt{2}}(\bs{E}_{k(n^{\prime}),\ell(n^{\prime})}+\bs{E}_{k(n^{\prime}),\ell(n^{\prime})}), & n=N+1, \dots, \frac{(N+1)N}{2}, & n^\prime = n-N \\
    \frac{1}{\sqrt{2}}(\bs{E}_{k(n^{\prime\prime}),\ell(n^{\prime\prime})}-\bs{E}_{k(n^{\prime\prime}),\ell(n^{\prime\prime})}), & n=\frac{(N+1)N}{2}+1, \dots, N^2, & n^{\prime\prime} = n-\frac{(N+1)N}{2}
    \end{matrix}\right.
\end{equation}
\end{figure*}

Now that we have defined a basis for $\mathfrak{u}(N)$ it only remains to define the transformation that allows extracting the $N^2$ real coordinates associated to each element $\bs{X}\in \mathfrak{u}(N)$.

\subsection{Unitary change of basis}\label{sec:unit_basis}
In Section~\ref{ss:exp_m}, we argued that $\mathfrak{u}(N)$ corresponds to a real vector space from its definition as a Lie algebra. As for any vector space, a change of basis can then be identified with a linear transformation, i.e., a matrix multiplication. Given the basis $\mathcal{B}$ defined in Section~\ref{ss:basis}, whose elements are given in \eqref{eq:basis}, we can write for each $\bs{X}\in \mathfrak{u}(N)$ 
\begin{equation}\label{eq:skew_basis}
    \bs{X} = \alpha_1 \bs{B}_1+\dots+\alpha_{N^2} \bs{B}_{N^2},
\end{equation}
where $\bs{\alpha}=[\alpha_1,\dots,\alpha_{N^2}]^\mrm{T}$ is the vector of real coordinates for $\bs{X}$. By vectorizing \eqref{eq:skew_basis} we can rewrite it as
\begin{equation}\label{eq:skew_basis_vec}
 \vec(\bs{X}) = \bs{U}_{\mathcal{B}} \bs{\alpha},
\end{equation}
where $\bs{U}_{\mathcal{B}}=[\vec(\bs{B}_1),\dots,\vec(\bs{B}_{N^2})]$. Since the Hilbert-Schmidt inner product coincides with the dot product of the vectorized input matrices, i.e., $\mrm{trace}(\bs{A}^\mrm{H}\bs{B})=\vec(\bs{A})^\mrm{H}\vec(\bs{B})$, the columns $\{\vec(\bs{B}_i)\}_{1\leq i\leq N^2}$ of $\bs{U}_{\mathcal{B}}$ are also orthonormal in the usual sense.\footnote{Note that the dot product corresponds to the standard inner product for complex vector spaces.} Hence, the $N^2\times N^2$ matrix $\bs{U}_{\mathcal{B}}$ corresponds itself to a unitary transformation, i.e., $\bs{U}_{\mathcal{B}}^\mrm{H}\bs{U}_{\mathcal{B}}=\bs{U}_{\mathcal{B}}\bs{U}_{\mathcal{B}}^\mrm{H}=\mathbf{I}_{N^2}$. The $N^2$ real coordinates associated to a given $\bs{X}\in \mathfrak{u}(N)$ can be thus obtained as
\begin{equation}\label{eq:basis_ch}
\bs{\alpha} = \bs{U}_{\mathcal{B}}^\mrm{H}  \vec(\bs{X}).
\end{equation}
The unitary property of the considered change of basis will be helpful in the forthcoming analysis. However, if the basis $\mathcal{B}$ was defined in any other way such that orthonormality is not achieved, the change of basis in \eqref{eq:basis_ch} would be applied by multiplying the inverse of $\bs{U}_{\mathcal{B}}$ instead of its conjugate transpose. This may lead to higher computational complexity due to the matrix inversion, which is not required if $\bs{U}_{\mathcal{B}}$ is unitary. However, since the same basis should be considered in any system employing this parametrization, said matrix inversion could be precomputed and reused indefinitely, incurring negligible complexity increase in the long run.

\begin{rmk}
    In some practical applications, e.g., in \ac{MIMO} CSI feedback, or in quantum state characterization, a unimodular scaling of the employed unitary matrix $\bs{U}$ may have no major impact on performance. For such cases, the first basis element in \eqref{eq:basis} may be ignored, and the resulting \ac{DEP} would be reduced to $N^2-1$ dimensions. This is equivalent to parameterize instead the projection of $\bs{U}$ to the special unitary group $\mathcal{SU}(N)$, for which \acp{DEP} have been studied in \cite{gen_gell_mann}. 
    Nevertheless, general applications, as the ones in Section~\ref{ssec:BD-RIS} and \ref{ssec:dec}, may still require to resolve this uncertainty, preventing any further dimensionality reduction. Moreover, for large enough $N$, the resulting dimensionality reduction from projecting onto $\mathcal{SU}(N)$ would be negligible.
\end{rmk}
\begin{rmk}
    In some applications we may need to share orthogonal matrices instead of unitary matrices, e.g., when sharing graph Fourier transform matrices as in \cite{GFT_sharing}. If we consider only orthogonal matrices corresponding to rotation matrices (having determinant 1), the \ac{DEP} may be achieved in the same way as the proposed unitary matrix \ac{DEP}, but removing the first $N(N+1)/2$ basis elements in \eqref{eq:basis}, associated to purely imaginary entries. This would reduce the resulting dimensions to $N(N-1)/2$ real dimensions. On the other hand, general orthogonal matrices (with determinant $\pm 1$) may be projected to rotation matrices by scaling the last row with the determinant. This would only require sending and extra bit associated to the respective determinant apart from the rotation matrix \ac{DEP}.
\end{rmk}

\subsection{Full parametrization and retrieval of unitary matrices}
If we combine the injective inverse exponential map, and the orthonormal basis inversion previously mentioned, we can define the \ac{DEP} described in \eqref{eq:fun_UR}, which allows parameterizing an arbitrary unitary matrix through a set of $N^2$ real numbers. Specifically, we can define $\bs{f}$ for each $\bs{U}\in \mathcal{U}(N)$ as
\begin{equation}\label{eq:f_final}
    \bs{f}(\bs{U}) = \bs{U}_{\mathcal{B}}^\mrm{H} \vec\big(\log(\bs{U})\big).
\end{equation}
On the other hand, the injectivity of $\bs{f}$ in ensured by the fact that $\log$ is injective, as previously mentioned, while the linear map associated to the change of basis $\bs{U}_{\mathcal{B}}^\mrm{H}$ is bijective since it corresponds to a full-rank linear transformation.

Assume that we have an arbitrary unitary matrix $\bs{U}\in \mathcal{U}(N)$ which was perfectly (without errors) stored/received using its \ac{DEP} $\bs{\alpha}=\bs{f}(\bs{U})$. We can then retrieve the original matrix $\bs{U}$ from its inverse map $\bs{U}=\bs{f}^{-1}(\bs{\alpha})$. Considering \eqref{eq:f_final}, we may define the inverse map as
\begin{equation}\label{eq:finv_final}
    \bs{f}^{-1}(\bs{\alpha}) = \exp \big(\vec^{-1}(\bs{U}_{\mathcal{B}}\bs{\alpha})\big),
\end{equation}
where $\vec^{-1}(\cdot)$ corresponds to inverse vectorization, which reorganizes the vector input into an $N\times N$ matrix. By restricting the domain of $\bs{f}^{-1}$ to the image $\bs{f}(\mathcal{U}(N))$ of the whole space of unitary matrices under $\bs{f}$, we may further ensure bijectivity so that the inverse corresponds to a well-defined function. However, the explicit definitions of the complex matrix exponential and logarithm from \eqref{eq:exp_skh} and \eqref{eq:log_unit}, respectively, allow us to relax this formality. The reason is that \eqref{eq:finv_final} will always retrieve the original $\bs{U}$ regardless of the considered image of the complex logarithm (for a similar reason that $\exp(\mrm{j}(x+\ell 2\pi))=\exp(\mrm{j}x)$, with $\ell\in\mathbb{Z}$, in the scalar case). This fact may be further understood after discussing boundedness in Section~\ref{sec:comp}.

\subsection{Time-complexity analysis}
Getting timely access to a unitary matrix from its \ac{DEP}, and viceversa, is of utmost importance for delay sensitive applications, as the ones considered in Section~\ref{section:app}. The following proposition determines the time-complexity scaling of the proposed \ac{DEP}.
\begin{prop}
    Assuming unlimited hardware resources, the time-complexity for computing the proposed \ac{DEP} given in \eqref{eq:f_final}, as well as its inverse map given in \eqref{eq:finv_final}, is $\mathcal{O}(N^3).$ 
\begin{proof}
    Let us start considering \eqref{eq:f_final}. The required operations consist of a matrix multiplication between an $N^2\times N^2$ matrix and an $N^2$ vector, as well as a matrix logarithm of the form \eqref{eq:log_unit}. The total number of operations to perform the matrix multiplication $\bs{U}_\mathcal{B}^\mrm{H}\vec(\bs{X})$ is $\mathcal{O}(N^4)$. However, if we have enough hardware resources, we may parallelize it by performing simultaneously (e.g., in different hardware modules) the multiplication of each element of the vector $\vec(\bs{X})$ with all the elements of the respective column of $\bs{U}_\mathcal{B}$, leading to time-complexity $\mathcal{O}(1)$. The $N^2$ sums of the resulting $N^2$-sized vectors can also be parallelized by summing simultaneously different entries of each vector, leading to time-complexity $\mathcal{O}(N^2)$. Thus, the time complexity for the matrix multiplication is $\mathcal{O}(N^2)$. For the matrix logarithm, the operations required are an \ac{EVD} with $\mathcal{O}(N^3)$, a logarithm of the $N$ eigenvalues with $\mathcal{O}(N)$, and two matrix multiplications: one $\mathcal{O}(1)$ (assuming parallelization) of an $N\times N$ diagonal matrix with a $N\times N$ matrix, and one $\mathcal{O}(N^3)$ of two $N \times N$ matrices. Hence, the overall time-complexity is $\mathcal{O}(N^3)$, which is dominated by the \ac{EVD} (and matrix multiplication) to perform \eqref{eq:log_unit}.

    For the inverse map \eqref{eq:finv_final}, the required operations are essentially the same as for \eqref{eq:f_final}, but taken in reverse order. The only difference is that the logarithm of the $N$ diagonal elements now corresponds to an exponential with the same complexity. Hence, the overall time-complexity is $\mathcal{O}(N^3)$, which is dominated by the \ac{EVD} (and matrix multiplication) to perform  \eqref{eq:exp_skh}.
\end{proof}
\end{prop}

It should be noted that most of the applications of interest for this work require at least pre-computing the \ac{SVD} of an $L\times N$ matrix, which has complexity $\mathcal{O}(\max(L,N)LN)$. Hence, for $L\geq N$, computing the proposed \ac{DEP} (or its inverse) would not increase the overall time-complexity order.
On the other hand, if we compare the complexity of the proposed \ac{DEP} with approaches based on the Cayley transform, as proposed in \cite{cayley_codes}, the time-complexity order would also be $\mathcal{O}(N^3)$ since this transform requires inverting an $N\times N$ matrix.

\section{Boundedness of the Dimensionally-Efficient Parametrization}\label{sec:comp}
In the previous section, we explicitly defined an injective function $\bs{f}$ that maps an arbitrary unitary matrix $\bs{U}\in \mathcal{U}(N)$ to its \ac{DEP} $\bs{f}(\bs{U})\in \mathbb{R}^{N^2}$. Next, we will show that $\bs{f}$ may be fixed such that the \ac{DEP} maintains the boundedness property, inherent to the unitary matrix space $\mathcal{U}(N)$. Specifically, we will show that the image of $\mathcal{U}(N)$ under $\bs{f}$ may be contained within a compact (i.e., closed and bounded) subspace of the form $[a,b]^{N^2}\subset\mathbb{R}^{N^2}$, so that we achieve \eqref{eq:fun_comp}.

Let us consider the inverse exponential map defined in \eqref{eq:log_unit} for an arbitrary $\bs{U}\in \mathcal{U}(N)$. As previously noted, the eigenvalues of a unitary matrix always lie on the complex unit circle \cite[Section 4.5]{unit_prop}, \cite{eig_unit_herm}. We may thus express
\begin{equation}
    \bs{\Lambda}_{\bs{U}}=\diag\big(\exp(\mrm{j}\varphi_1),\dots,\exp(\mrm{j}\varphi_N)\big),
\end{equation}
where $\varphi_n\in \mathbb{R}$, $\forall{n}$. From the periodicity of the complex exponential, we may further consider the common restriction $\varphi_n\in (-\pi,\pi]$, $\forall{n}$. If we then apply the complex logarithm to the diagonal elements, we get
\begin{equation}\label{eq:eig_skew}
    \log(\bs{\Lambda}_{\bs{U}})=\diag \big(\mrm{j}(\varphi_1+\ell_1 2\pi),\dots,\mrm{j}(\varphi_N+\ell_N 2\pi)\big),
\end{equation}
where $\ell_n\in\mathbb{Z}$, $\forall{n}$. Note that the space of Skew-Hermitian matrices (correspondingly $\mathfrak{u}(N)$) coincides with the space of normal matrices with purely imaginary eigenvalues since, for $\bs{X}\in \mathfrak{u}(N)$, we have
\begin{equation}
    \bs{X}+\bs{X}^\mrm{H}=\bs{U}_{\bs{X}} (\bs{\Lambda}_{\bs{X}}+\bs{\Lambda}_{\bs{X}}^{*})\bs{U}_{\bs{X}}^\mrm{H},
\end{equation}
which equals $\bs{0}$ if and  only if $\bs{\Lambda}_{\bs{X}}=-\bs{\Lambda}_{\bs{X}}^{*}$, i.e., if and only if the eigenvalues are purely imaginary. Thus, we could in principle span the whole $\mathfrak{u}(N)$ by considering the complete logarithm definition from \eqref{eq:eig_skew} in \eqref{eq:log_unit} for each $\bs{U}\in\mathcal{U}(N)$. However, given that the idea is to eventually go back to the unitary group, we may restrict without loss the definition of the complex logarithm to its principal brach, such that the outputs lie always in the interval $(-\pi,\pi]$ of the imaginary axis. This corresponds to considering a logarithm definition as \eqref{eq:eig_skew}, but where we fix $\ell_n=0$,  $\forall{n}$. In the remainder, we assume that any system employing the proposed unitary matrix \ac{DEP} considers the latter definition such that, $\forall \bs{U}\in\mathcal{U}(N)$, the eigenvalues of $\log(\bs{U})$ may be expressed as $\lambda_n \big(\log(\bs{U})\big)=\mrm{j}\varphi_n$, with $\varphi_n\in(-\pi,\pi]$, $\forall n$. Note that this definition also allows to have a well defined logarithm for unitary matrices $\bs{U}$ having some eigenvalues equal to $-1$, which would give $\varphi_n=\pi$.

The following proposition stresses that the proposed unitary matrix \ac{DEP} can ensure boundedness, while it provides a specific bound on the resulting dynamic range.
\begin{prop}\label{prop:comp}
Assume $\bs{f}$ is defined by \eqref{eq:f_final}, where $\log$ is defined by \eqref{eq:log_unit} and \eqref{eq:eig_skew} with $\ell_n =0$, $\forall n$.  The image of $\mathcal{U}(N)$ under $\bs{f}$ is contained within the compact (bounded) space ${\big[-\sqrt{N}\pi,\sqrt{N}\pi\big]^{N^2}\subset \mathbb{R}^{N^2}}$. In other words, $\forall \bs{U}\in \mathcal{U}(N)$ we have that the entries of $\bs{f}(\bs{U})$ fulfill 
\begin{equation}\label{eq:f_i_int}
    f_i(\bs{U})\in\big[-\sqrt{N}\pi,\sqrt{N}\pi\big], \;\;\;\;\forall i \in \{1,\dots,N^2\}.
\end{equation}
\begin{proof}
Using the logarithm definition from \eqref{eq:eig_skew} with $l_n=0$, $\forall n$, we may express \eqref{eq:f_final} for any $\bs{U}\in \mathcal{U}(N)$ as
\begin{equation}\label{eq:f_diag}
\begin{aligned}
    \bs{f}(\bs{U}) &= \bs{U}_{\mathcal{B}}^\mrm{H} \vec\big(\bs{U}_{\bs{U}} \diag(\mrm{j}\varphi_1,\dots,\mrm{j}\varphi_N)\bs{U}_{\bs{U}}^\mrm{H} \big)\\
    &= \bs{U}_{\mathcal{B}}^\mrm{H}(\bs{U}_{\bs{U}}^* \otimes \bs{U}_{\bs{U}})\vec\big(\diag(\mrm{j}\varphi_1,\dots,\mrm{j}\varphi_N)\big).
\end{aligned}
\end{equation}
Given the orthonormal basis defined in Section~\ref{sec:unit_basis}, the product 
$\bs{U}_{\mathcal{B}}^\mrm{H}(\bs{U}_{\bs{U}}^* \otimes \bs{U}_{\bs{U}})$ corresponds to a product of unitary matrices, giving itself an $N^2 \times N^2$ unitary matrix. Hence, the squared norm of \eqref{eq:f_diag} is given by
\begin{equation}\label{eq:norm_f}
    \Vert\bs{f}(\bs{U})\Vert^2\triangleq \bs{f}^{\mrm{H}}(\bs{U})\bs{f}(\bs{U})=\sum_{n=1}^{N}\varphi_n^2.
\end{equation} 
Note that, although not explicitly seen from \eqref{eq:f_diag}, the selected basis derived in Section~\ref{sec:unit_basis} ensures that each entry, $f_i(\bs{U})$, of $\bs{f}(\bs{U})$ is real, $\forall i\in\{1,\dots,N^2\}$. Since the considered logarithm definition further ensures that $\varphi_n\in (-\pi,\pi]$, we can upper-bound \eqref{eq:norm_f} as
\begin{equation}
    \Vert \bs{f}(\bs{U})\Vert^2\triangleq \sum_{i=1}^{N^2}f_i^2(\bs{U})\leq N\pi^2,
\end{equation}
which trivially leads to
\begin{equation}\label{eq:upper_bd}
    \vert f_i(\bs{U})\vert \leq \sqrt{N}\pi, \;\;\;\;\forall i\in\{1,\dots,N^2\}.
\end{equation}
The proof finalizes by noting that \eqref{eq:upper_bd} is equivalent to \eqref{eq:f_i_int}.

\end{proof}
\end{prop}
\if
Hence, \eqref{eq:f_diag} corresponds to a combination of $N$ columns from a unitary matrix, scaled by the purely complex eigenvalues $\mrm{j}\varphi_n$, $n=1,\dots, N$, i.e.,
\begin{equation}\label{eq:comb_}
    \bs{f}(\bs{U}) = \sum_{n=1}^{N}\mrm{j}\varphi_n \bs{q}_{n},
\end{equation}
with $\Vert\bs{q}_{n}\Vert^2=1$ and $\bs{q}_{i}^\mrm{H}\bs{q}_j=0$ for $i\neq j$.
\fi

With the results presented in this section, we can now proclaim that we have explicitly specified an injective function
\begin{equation}\label{eq:fun_comp_fin}
    \bs{f}: \mathcal{U}(N)\to \left[-\sqrt{N}\pi,\sqrt{N}\pi\right]^{N^2},
\end{equation}
as well as its inverse map. Thus, we can use these maps to unequivocally parameterize unitary matrices as a set of $N^2$ real bounded numbers. This may allow reducing the amount of information required to identify an arbitrary unitary matrix with respect to naively identifying it through its $N^2$ complex entries (i.e., $2N^2$ real entries). Moreover, the boundedness property also allows for the use of effective quantization with bounded error. Note, however, that the real and imaginary parts of the entries of an $N\times N$ unitary matrix are also bounded since they are contained within the interval $[-1,1]$. Next, we emphasize the value of the proposed \ac{DEP} by presenting several applications from the field of wireless communications which may benefit from it.

\section{Wireless Communications Applications}\label{section:app}
One of the most relevant modern advancements in the field of wireless communications is the use of \ac{MIMO} \cite{mimo}, which can significantly improve spectral efficiency by exploiting the degrees of freedom in the spatial domain. Much of the \ac{MIMO} processing depends on performing operations involving unitary matrices, e.g., diagonalization of the channel matrix. These unitary matrices should be stored, and sometimes even shared among different modules. Next, we specify three potential applications related to the field of \ac{MIMO} communications that showcase the benefit of the unitary matrix \ac{DEP} proposed in this work. However, as previously outlined, the scope of this work may go beyond these applications, finding potential utility within diverse fields as image processing, quantum computation, computer science, etc.

\subsection{CSI-Feedback to Achieve MIMO Capacity}\label{ssec:CSI}
Perhaps the most direct application of the proposed \ac{DEP} is within \ac{MIMO} \ac{CSI} feedback, which is a topic that has been widely studied in the literature \cite{givens_CSI,givens2_spatial_mat,stiefel_pred_quant,diff2}. Let us consider an uplink transmission of a $N$-antenna \ac{UE} towards an $M$-antenna \ac{BS} ($M\geq N$) under a narrowband channel. The received vector at the \ac{BS} is then governed by the renowned \ac{MIMO} equation \cite{mimo}
\begin{equation}\label{eq:mimo_eq}
\bs{y}=\bs{H}\bs{s}+\bs{n},
\end{equation}
where $\bs{H}$ is the $M\times N$ complex channel matrix, $\bs{s}$ is the vector of baseband symbols transmitted by each of the \ac{UE}'s antennas, and $\bs{n}\sim\mathcal{N}_{\mathbb{C}}(\bs{0},\mathbf{I}_M)$ is the noise vector. In order to achieve the channel capacity, the \ac{UE} should access the orthogonal spatial streams by precoding its data according to the right unitary matrix of the \ac{SVD} of the channel, which allows subsequently allocating power to each spatial stream according to the waterfilling algorithm \cite{telatar}. However, in modern \ac{MIMO}-based systems, the channel is typically assumed to be estimated via uplink pilots \cite{mMIMObook}, which means that only the \ac{BS} has access to the estimate of $\bs{H}$. Thus, a CSI-feedback strategy should be employed so that the \ac{UE} can gain access to the relevant \ac{CSI} required to achieve capacity.

Assuming that the \ac{BS} performs the \ac{SVD} of the channel matrix, given by $\bs{H}=\bs{U}\bs{\Sigma}\bs{V}^\mrm{H}$,\footnote{Note that performing this \ac{SVD} at the \ac{BS} is needed to be able to apply the linear equalizer $\bs{U}^\mrm{H}$ that allows achieving capacity \cite{mimo}.} the proposed \ac{DEP} could be employed to reduce the information rate required to feedback the matrix $\bs{V}$ to the \ac{UE}. In order to achieve capacity, we should also send the $N$ real numbers associated to the power allocation, so that the total number of real dimensions to be feedbacked to the \ac{UE} would be $N^2+N$. Since the power allocation values may be given as ratios with respect to the total available power, the whole $N^2+N$ real dimensions are associated to bounded quantities, which can be effectively quantized with limited reconstruction error. Thus, for large enough $N$, we could essentially halve the dimension of the information required to feedback these quantities with respect to sending the complete $\bs{V}$ matrix together with the power allocation values, giving instead $2N^2+N$ bounded real dimensions. Another feedback option with comparable dimensional-efficiency could be to send the Gramian of the channel, upon which the \ac{UE} would have to perform \ac{SVD} independently, leading to the same time-complexity order as for retrieving the respective unitary matrix from the considered \ac{DEP}. However, the elements of the Gramian matrix are not bounded in general, which may lead to greater quantization errors. Moreover, the \ac{UE} would also have to find the optimum power allocations from the eigenvalues of the received Gramian, leading to higher power consumption at the device, which may compromise its battery life.

\subsection{Fully-Connected RIS Configurations}\label{ssec:FRIS}
\label{ssec:BD-RIS}
RIS is a wireless communication technology which has gained much attention in the context 6G and beyond research. Typical \ac{RIS} implementations consist of a surface equipped with a large number of passive reconfigurable elements, which may change their reflection properties to improve wireless communication links. We may consider a \ac{MU-MIMO} scenario where an $M$-antenna \ac{BS} serves $K$ single-antenna \acp{UE}. The uplink system model may be described through an input-output relation as \eqref{eq:mimo_eq}, but where the channel matrix is given by
\begin{equation}
    \bs{H} = \bs{H}_0+\bs{H}_1\bs{\Theta}\bs{H}_2,
\end{equation}
where $\bs{H}_0$ is the $M\times K$ direct channel matrix between the \acp{UE} and the \ac{BS}, $\bs{H}_1$ is the $M\times N$ channel matrix between the \ac{BS} and the \ac{RIS}, $\bs{\Theta}$ is the $N\times N$ reconfigurable reflection matrix, and $\bs{H}_2$ is the $N\times K$ channel matrix between the \acp{UE} and the \ac{RIS}.

Recent works have considered extended \ac{RIS} architectures by assuming that the \ac{RIS} elements can be interconnected, offering a greater degree of reconfigurability \cite{bd-ris,bd-ris1}. These architectures fall within the umbrella of \ac{BD-RIS} since the associated reflection matrix $\bs{\Theta}$ may depart from the common diagonal reflection matrix with unimodular entries considered in the conventional \ac{RIS}. In fact, if we consider fully-connected \ac{BD-RIS} implementations with lossless passive impedance networks, the reflection matrix becomes unitary \cite{bd-ris,bd-ris1}. Due to the passive nature of these surfaces, when they are employed in \ac{MU-MIMO} scenarios, the \ac{BS} is typically in charge of estimating the channel and finding the appropriate \ac{BD-RIS} configuration \cite{RS_orth,bd-ris_ch_est}, which can be unequivocally identified with its reflection matrix. Thus, the \ac{BS} should then employ a side-link to forward the respective configuration to the \ac{BD-RIS} so that it can apply it. Assuming a \ac{BD-RIS} with full-reconfigurability, and exploiting the unitary constraint on the reflection matrix, the proposed unitary matrix \ac{DEP} may be employed to reduce the overhead required to forward the respective configuration from the \ac{BS} (or generic receiver) to the \ac{BD-RIS}.

\begin{rmk} \label{rmk:rec_bdris}
The proposed \ac{DEP} is especially useful when considering a fully-connected lossless non-reciprocal \ac{BD-RIS} \cite{nr-BD-RIS}, also found under the term \ac{FRIS} \cite{RS_orth}. The reason is that such system would pose no further constraints on the reflection matrix other than the unitary constraint. However, the proposed \ac{DEP} could be adapted to take into account symmetry constraints (associated to reciprocal \ac{BD-RIS}) to further reduce dimensionality. This may be simply done by eliminating some of the basis elements in \eqref{eq:basis}. Specifically, it is enough to remove the $N(N-1)/2$ real anti-symmetric elements, since the matrix logarithm of a symmetric unitary matrix only has purely imaginary entries.\footnote{Note that if, $\bs{U}=\bs{U}^\mrm{T}$, the eigenvalue decomposition in \eqref{eq:log_unit} has real eigenvector matrix since $\bs{U}_{\bs{U}}=\bs{U}_{\bs{U}}^*$, while the resulting eigenvalues are purely imaginary.} The resulting \ac{DEP} would then be reduced to $N(N+1)/2$ bounded real numbers. On the other hand, in case of further restrictions on the interconnection structure of the \ac{BD-RIS}, e.g., leading to a block diagonal reflection matrix, the dimensionality of the proposed \ac{DEP} may be similarly reduced by removing other unnecessary basis elements from \eqref{eq:basis}.
\end{rmk}

An alternative to sending the unitary matrix associated to the \ac{BD-RIS} reflection matrix is to send the impedance values associated to the configuration leading to such reflection matrix. If we consider lossless non-reciprocal \ac{BD-RIS} implementations based on circulators, as illustrated in \cite{RS_orth} for \ac{FRIS}, the respective impedance matrix may be linked to $2N^2+N$ real numbers (scaled by the imaginary unit), which would thus require more than twice the dimensions of the proposed \ac{DEP}. For the case of reciprocal \ac{BD-RIS}, the impedance matrix would correspond to a symmetric purely imaginary matrix \cite{bd-ris1}, giving $N(N+1)/2$ real variables as in the proposed \ac{DEP} after removing the unnecessary basis elements from \eqref{eq:basis}. However, due to the relation between the reflection matrix and the impedance matrix, which is given through the Cayley transform \cite{pozar}, the resulting impedance values would not fulfill the boundedness property. This would make them less suitable for transmission since they could lead to unbounded quantization errors, as previously noted.

\subsection{Sharing of Decentralized Analog Beamformers}\label{ssec:WAX}
\label{ssec:dec}
As happens with the reflection matrix of the lossless impedance networks used in BD-RIS architectures, the narrowband response of a passive and lossless analog beamformer can be identified with a unitary matrix associated to the corresponding impedance network \cite{pozar}. An example of such is the Butler matrix, proposed as an analog beamformer based on phase shifters and power dividers \cite{butler_matrix}. In \cite{asilomar24}, a generalized hybrid beamforming architecture is considered where the analog beaforming may be performed in decetralized modules. The resulting processing is described through the WAX framework \cite{wax_journal, A_struct_journal}, which allows exploiting the trade-off between decentralized processing complexity and level of decentralization. The post-processed vector is then given by
\begin{equation}
    \bs{z} = \bs{X}^\mrm{H}\bs{A}^\mrm{H}\bs{W}^\mrm{H}\bs{y},
\end{equation}
where $\bs{y}$ is the received vector (as in \eqref{eq:mimo_eq}), $\bs{X}$ is the baseband processing, $\bs{A}$ is a fixed analog combining module, and $\bs{W}=\diag(\bs{W}_1,\dots,\bs{W}_{M_\mrm{P}})$ is the analog beamforming matrix, where each $\bs{W}_m$ corresponds to an $N\times N$ unitary matrix associated to the analog beamforming applied at decentralized module $m$. This general scheme, and other related ones, potentially rely on finding the analog beamformers at a central node, associated to a \ac{BBU}, which should then be forwarded to the decentralized modules where they are physically applied. Alternatively, these analog beamformers may be found iteratively by updating and sharing them among the different decentralized modules. The interconnection bandwidth required to forward the unitary matrices (associated to the analog beamformers) from the central node to the decentralized nodes, or to iteratively share them among the different decentralized modules, may be effectively reduced by employing the proposed unitary matrix \ac{DEP}.

\section{Numerical Results}
\subsection{Analysis of Reconstruction Error}
To understand the potential of the proposed \ac{DEP} for efficient transmission/storage of unitary matrices, we begin by analyzing the error incurred when retrieving the unitary matrices from noisy observations of the \ac{DEP}. A common metric for measuring the error in a reconstructed matrix is the \ac{MSE}, which may be computed using the matrix Frobenius norm as
\begin{equation}\label{eq:MSE_def}
    \mrm{MSE} = \frac{1}{N^2}\mathbb{E}\{\Vert \bs{U}-\widehat{\bs{U}}\Vert_{\mrm{F}}^2\}.
\end{equation}
where the expectation averages throughout different realizations of the original matrix $\bs{U}$ and its reconstructed version $\widehat{\bs{U}}$, while, for each realization, the error is averaged throughout the $N^2$ matrix entries. Note that, although \ac{MSE} is not directly tailored to unitary matrices, the squared distance induced by the standard Hilbert-Schmidt inner product for unitary matrices (or trace distance) has direct correspondence with the \ac{MSE} defined in \eqref{eq:MSE_def} through a scaling factor of $N/2$. Another way to analyze the reconstruction error is to look at the fidelity between unitary operators \cite{fidelity}, which is widely used in the context of quantum gates \cite{fidelity_use}. We consider the following normalized definition
\begin{equation}
    F(\bs{U},\widehat{\bs{U}}) = \frac{1}{N}\vert\mrm{Tr}(\bs{U}\widehat{\bs{U}})\vert,
\end{equation}
which gives a value between 0 and 1, where 1 corresponds to perfect reconstruction. Note that this definition of fidelity ignores the global phase uncertainty, further focusing on orthogonal space retrieval. For performance comparison, aside from the proposed \ac{DEP}, we will consider a naive approach, consisting of transmitting the full $2N^2$ real dimensions associated to the matrix entries, as well as an improved version where the resulting matrix is projected back to the space of unitary matrices.\footnote{Projecting an arbitrary matrix to $\mathcal{U}(N)$ requires computing the \ac{SVD}, leading to the same time-complexity order required to retrieve a unitary matrix from the proposed \ac{DEP}.} We will also consider a more advanced \ac{DEP} based on Givens rotations, similar to what is considered in \cite{givens_CSI,givens2_spatial_mat,GFT_sharing}, which may also be expressed in terms of $N^2$ bounded real dimensions: $N^2-N$ for the normalized Givens rotations, given by phases and amplitudes smaller than 1, and $N$ phase values to resolve the uncertainty associated to multiplying by a diagonal matrix of unimodular entries.

\begin{figure*}[h]
\begin{subfigure}{\columnwidth}
\centering
    \includegraphics[scale=0.55]{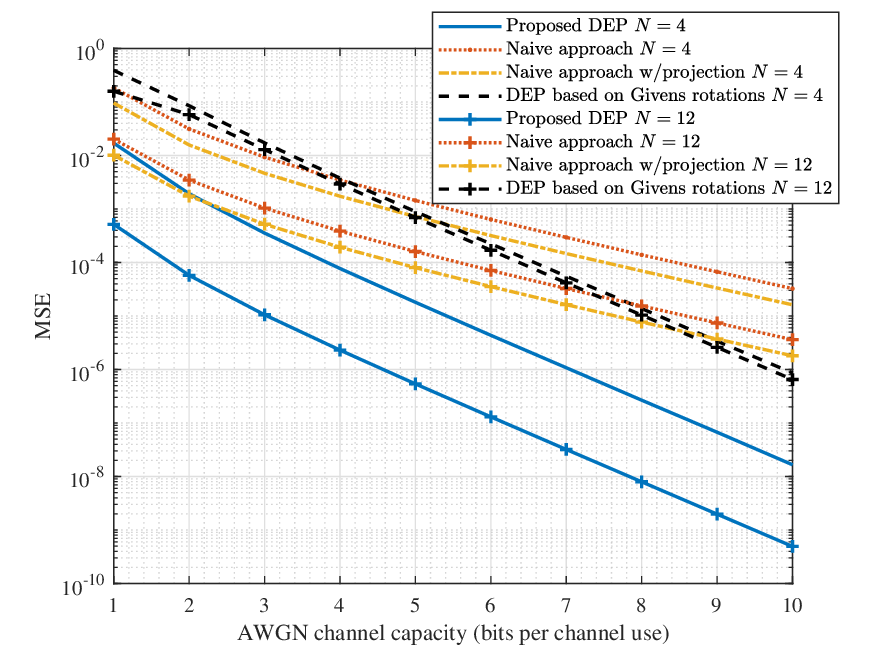}
  \end{subfigure}
  \begin{subfigure}{\columnwidth}
  \centering
    \includegraphics[scale=0.55]{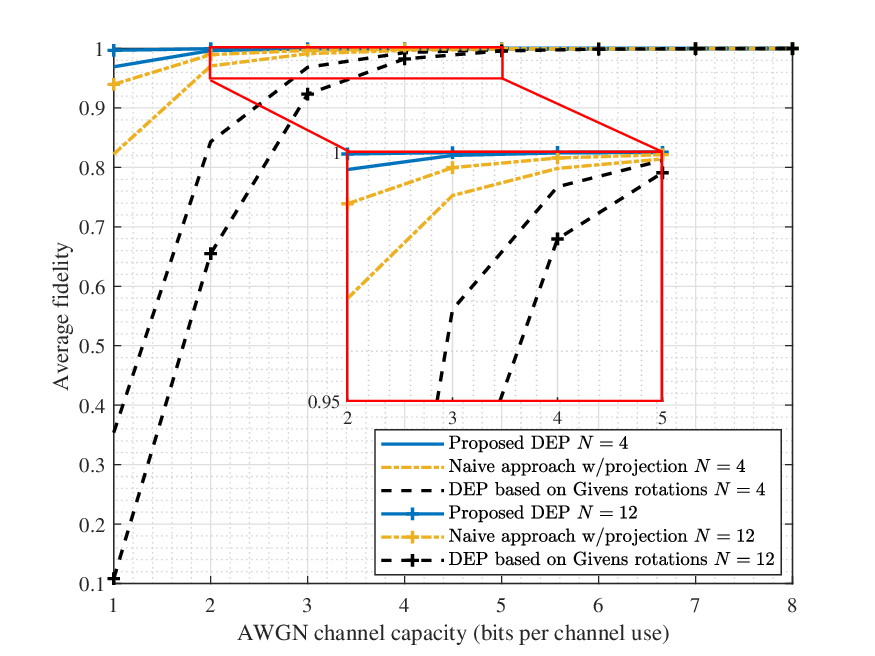}
\end{subfigure}
  \caption{\ac{MSE} (left) and average fidelity (right)  versus capacity per channel use for \ac{AWGN} channel.}
  \label{fig:AWGN}
\end{figure*}

In Fig.~\ref{fig:AWGN} we plot the \ac{MSE} (left) and average fidelity (right) when retrieving a unitary matrix from its noisy \ac{DEP} under Gaussian noise. We have averaged the results over $10^4$ realizations of an isotropic (uniformly distributed) $\bs{U}\in \mathcal{U}(N)$. The results assume that each \ac{DEP} entry suffers from \ac{AWGN}, where the capacity per channel use has full correspondence with the reconstruction \ac{SNR} through the renowned $\log_2(1+\mrm{SNR})$ equation \cite{inf_th}. For the \acp{DEP}, we consider $N^2$ channel uses (one per dimension), while for the naive approaches we consider $2N^2$ channel uses (one per real/imaginary matrix entry) with half the reported capacity, which corresponds to equally dividing the capacity between the the real and imaginary parts. It can be seen that the proposed \ac{DEP} offers the best reconstruction performance in terms of both \ac{MSE} and average fidelity. The \ac{DEP} based on Givens rotations gets the same \ac{MSE} slope as the proposed \ac{DEP} since they both achieve the same number of dimensions. However, the proposed \ac{DEP} attains a lower \ac{MSE} offset, especially as the number of dimensions increase. As for the naive approaches, including the projection back to $\mathcal{U}(N)$ reduces the reconstruction \ac{MSE} as expected, since the extra noise falling outside the unitary group is removed. However, the same slope is maintained with respect to the naive approach without projection due to the unchanged number of dimensions. In terms of Fidelity, the proposed \ac{DEP} converges quickly to full-fidelity, while the naive approach including projection attains a higher fidelity than the \ac{DEP} based on Givens rotations. Note that the definition of fidelity only applies to unitary matrices, while the naive approach without projection may give a non-unitary reconstructed matrix.



\begin{figure*}[h]
\begin{subfigure}{\columnwidth}
\centering
    \includegraphics[scale=0.55]{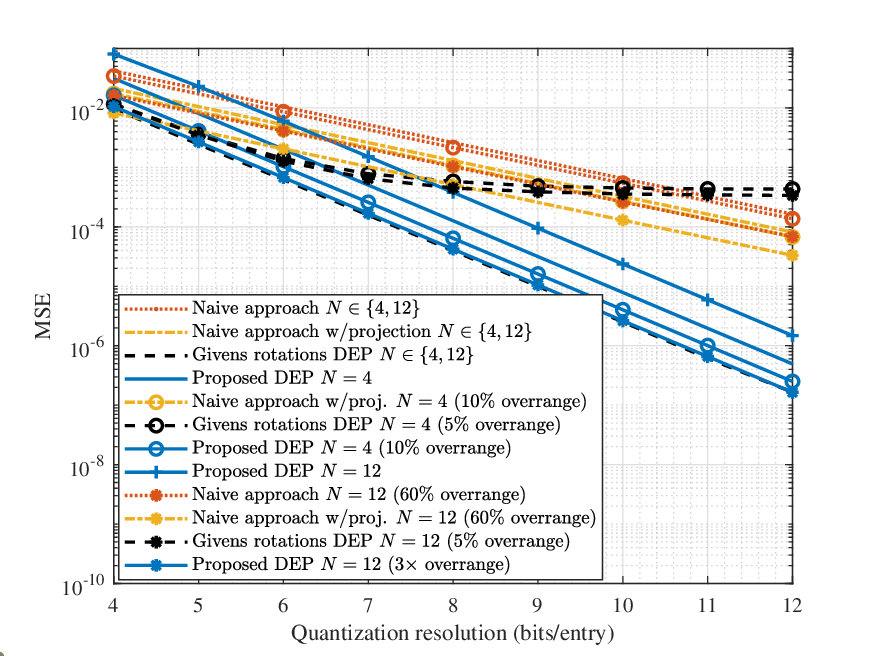}
  \end{subfigure}
  \begin{subfigure}{\columnwidth}
  \centering
    \includegraphics[scale=0.55]{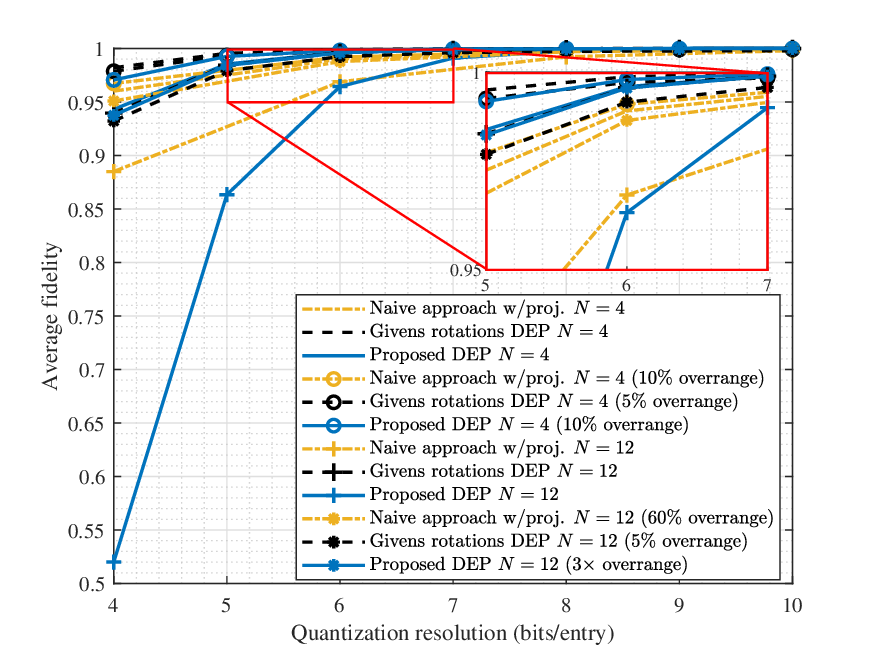}
\end{subfigure}
  \caption{Reconstruction error versus quantization resolution per entry.}
  \label{fig:quant}
  \vspace{-0.5em}
  \end{figure*}

In practical scenarios, the \ac{DEP} would be stored/transmitted in the digital domain using simple quantizers. The most basic and widespread way to perform quantization is uniform scalar quantization \cite{quant_surv}, which consists of dividing the range of input values into a discrete set of equally-sized intervals. Fig.~\ref{fig:quant} shows the \ac{MSE} (left) and average fidelity when retrieving $10^4$ isotropic unitary matrices from their uniformly quantized \ac{DEP}. The quantization resolution corresponds to the bits used to quantize each of the $N^2$ entries of the \acp{DEP}, while these bits are equally divided between the real and imaginary entries in the naive approaches. In all cases, the quantization ranges have been fixed to the known bounds of the input values, which were derived in Proposition~\ref{prop:comp} for the proposed \ac{DEP}, for the Givens rotations based \ac{DEP} some lie in the $[0,1]$ interval and some in the $(-\pi,\pi]$ interval \cite{givens_CSI}, and for the naive approaches all values are in the $[0,1]$ interval. We have also included results under \textit{quantization overrange}, which consist of allowing the quantizer inputs to have a dynamic range beyond the quantization range by a certain margin (in percentage or scaling). The input range for the proposed \ac{DEP} is assumed to coincide with \eqref{eq:f_i_int}, and symmetry around the midpoint is always assumed.

As  we can see from Fig.~\ref{fig:quant} (left), the proposed \ac{DEP} and the Givens rotations based one both attain the same \ac{MSE} slope, which is greater than for the naive approaches due to the reduced dimensions. However, the proposed \ac{DEP} has slightly worse performance than the Givens rotations based one, with greater loss as the matrix size increases. However, as we allow for some overrange, the proposed \ac{DEP} can attain similar performance to the Givens rotations based one, with more ground for improvement as the matrix size increases. Note that the interval derived in Proposition~\ref{prop:comp} for bounding the proposed \ac{DEP} is not necessarily tight, and more research may be needed on how to further reduce it. However, a reasonable explanation is that this may come from the fact that the proposed \ac{DEP} generates outputs that are more concentrated around the mean, while in the Givens based \ac{DEP} the entries may have a probability distribution closer to uniform. Specifically, from the way the boundedness interval is derived, if one value is close to the interval boundary, the rest of the entries will have to be very far from it. The same happens with the naive approaches, where, if one entry is close to 1, the rest of the entries associated to that column/row are likely to be close to 0 due to the norm constraint. Hence, as the number of entries grow, it is less likely to have entries close to the boundaries of the input range, allowing for some gains from quantization overrange. In the case of the Givens rotations based \ac{DEP}, it seems that the slight overrange (of 5$\%$) already starts to compromise performance, reducing the flexibility of this approach. Similar conclusions can be derived from the average fidelity metric in Fig.~\ref{fig:quant} (right), although in this case the Givens rotations based method seems less affected by overrange, while all approaches are negatively affected when increasing matrix size. In general, we can conclude that the proposed \ac{DEP} can offer good performance when allowing for some overrange, while its performance may benefit hugely when considering more advanced quantization techniques, e.g., non-uniform quantization, or even vector quantization. Analyzing the distribution of the proposed \ac{DEP} under various input distributions, and designing tailored quantization for it poses a research challenge in itself, which may be considered in future work. In fact, a simple adjustment of the uniform quantization taking into account estimates of the standard deviation of the different \ac{DEP} entries could greatly improve the performance, e.g., considering the "four-sigma" rule of thumb to find good overrange factors \cite{4std}.

\subsection{MIMO CSI-feedback capacity ratio}
We next analyze the performance of the considered approaches in the \ac{CSI}-feedback use case presented in Section~\ref{ssec:CSI}. We define as performance metric the achievable capacity ratio, which corresponds to the average ratio $\mathbb{E}_{\bs{H}}\{R_\mrm{ach}/C\}$ of the channel capacity $C$ (achievable through perfect waterfilling \cite{telatar}) and the achievable rate $R_\mrm{ach}$ when using the reconstructed unitary matrix for precoding at the \ac{UE}, together with the waterfilling power allocation. $R_\mrm{ach}$ may be easily computed as $\sum_{i}\log_2(1+\mrm{SINR}_i)$ from the resulting \ac{SINR} per stream, since the \ac{BS} assumes that the precoding is capable of orthogonalizing the streams. We have averaged the results throughout $10^3$ realizations of a standard \ac{IID} Rayleigh fading channel. We consider different number of \ac{UE} antennas, but the number of \ac{BS} antennas is fixed to $M=32$ antennas, and the communication \ac{SNR} is fixed to $10$ dB, since these have no direct impact on the considered approaches. With respect to the feedack channel characteristics, the same consideration apply as for Figs. \ref{fig:AWGN} and \ref{fig:quant} for both the \ac{AWGN} and the quantization cases. For the quantization case, we have further assumed that the waterfilling power allocation values are also quantized using the same resolution per entry.

\begin{figure*}[h]
\begin{subfigure}{\columnwidth}
    \centering
    \includegraphics[scale=0.55]{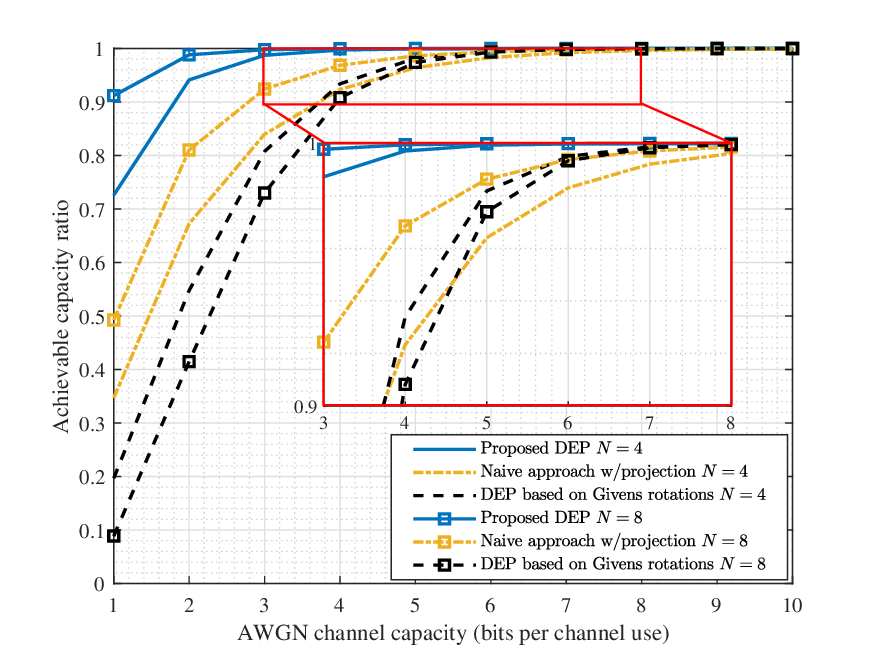}
  \caption{AWGN feedback channel.}
  \label{fig:AWGN_CSI}
\end{subfigure}
  \begin{subfigure}{\columnwidth}
  \centering
    \includegraphics[scale=0.55]{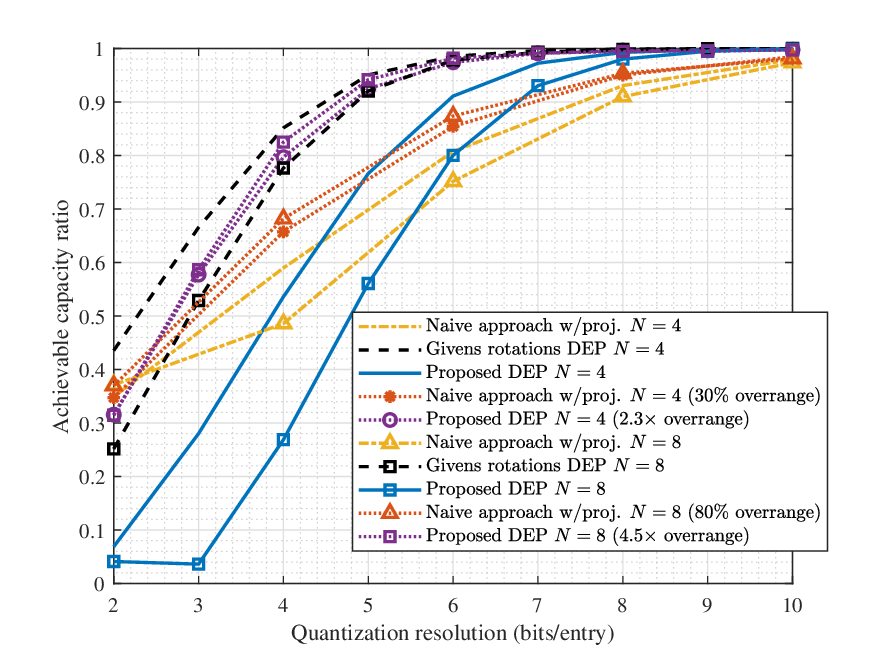}
  \caption{Quantized feedback.}
  \label{fig:q_CSI}
\end{subfigure}
\caption{Achievable capacity ratio versus quantization resolution for CSI feedback scenario with $M=32$ \ac{BS} antennas.}
\vspace{-0.5em}
\end{figure*}

In Fig.~\ref{fig:AWGN_CSI} we see that the proposed \ac{DEP} attains the best performance under \ac{AWGN} feedback channel, requiring only around 3-4 bit \ac{AWGN} channels to attain the maximum capacity. This is roughly half the feedback channel capacity required for both the naive approach, as well as the Givens rotations \ac{DEP}, to achieve maximum capacity. For the case with $N=4$ \ac{UE} antennas, the Givens based \ac{DEP} can only outperform the naive approach for large enough feedback channel capacity, in agreement with what was seen in Fig.~\ref{fig:AWGN}, while its performance is further degraded as the number of \ac{UE} antennas increases to $N=8$. In Fig.~\ref{fig:q_CSI} we see again that, under uniform quantization, the proposed \ac{DEP} can suffer from some performance loss with respect to the Givens rotations \ac{DEP}. However, as we allow for some quantization overrange, the proposed \ac{DEP} can even outperform the Givens rotations \ac{DEP} for large enough $N$. This hints again at the fact that the proposed approach can greatly benefit from more tailored quantization schemes, which may be an interesting direction for future work. We have not included any plots for the Givens rotations \ac{DEP} with quantization overrange to avoid cluttering the figure, since the slightest overrange already incurs some degradation in its performance.

\subsection{FRIS sum capacity ratio}
Next, we analyze the performance of the previous approaches in the fully-connected \ac{BD-RIS} use case described in Section~\ref{ssec:BD-RIS}. We consider a scenario with a \ac{FRIS}, or non-reciprocal \ac{BD-RIS}, where the direct link is blocked leading to $\bs{H}_0 = \bs{0}$, since in this scenario we have a closed-form expression for the \ac{FRIS} configuration achieving the sum capacity, as derived in \cite{bartoli}. However, the presented methods are also applicable to the case of having a direct channel since we consider the transmission of the full unitary configuration matrix. Moreover, the proposed \ac{DEP} can be trivially adapted to the reciprocal \ac{BD-RIS} scenario with symmetric reflection matrix, reducing the required overhead as described in Remark~\ref{rmk:rec_bdris}. The naive approach can also trivially incorporate this symmetry restriction, but for the Givens rotations \ac{DEP} this corresponds to finding a subset of Givens rotations whose product can give an arbitrary symmetric unitary matrix. This is a highly non-trivial problem, so the practicality of the Givens rotations \ac{DEP} when dealing with reciprocal \ac{BD-RIS} scenarios (or other physical restrictions) is questionable.

\begin{figure*}[h]
\begin{subfigure}{\columnwidth}
  \centering
    \includegraphics[scale=0.55]{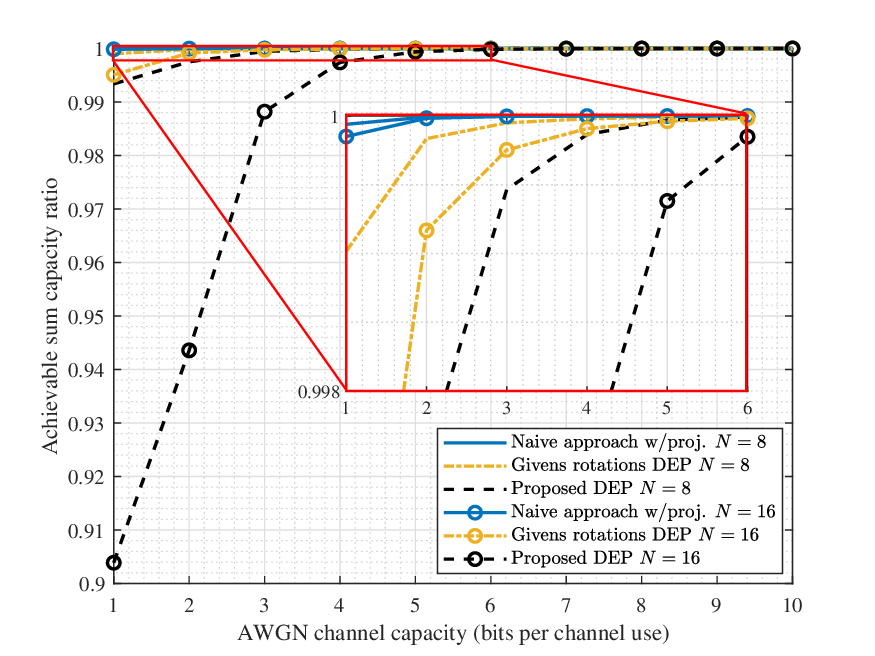}
  \caption{AWGN feedback channel.}
  \label{fig:AWGN_FRIS}
  \end{subfigure}
  \begin{subfigure}{\columnwidth}
  \centering
    \includegraphics[scale=0.55]{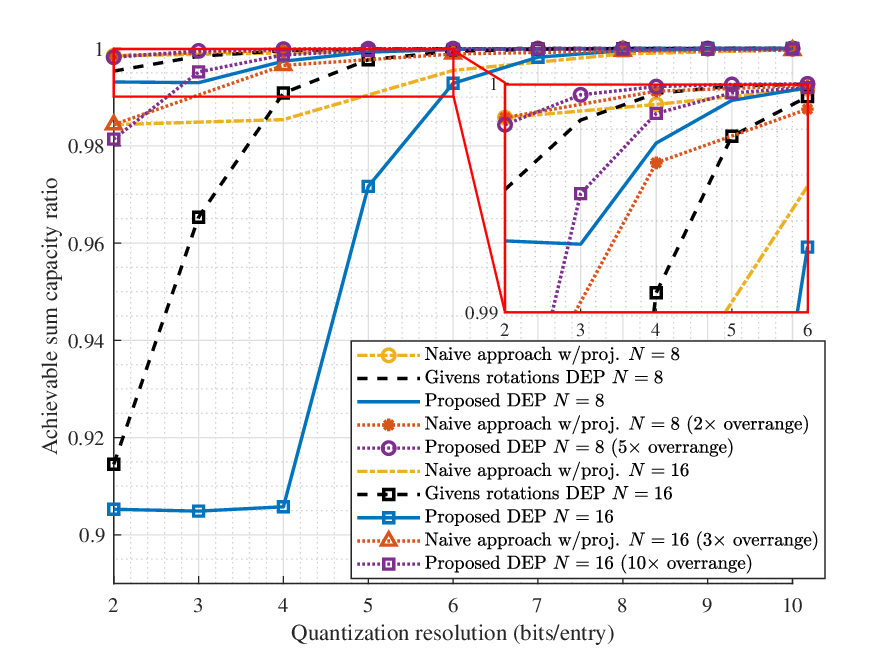}
  \caption{Quantized feedback.}
  \label{fig:q_FRIS}
\end{subfigure}
\caption{Achievable sum capacity ratio for FRIS control feedback scenario with $M=16$ \ac{BS} antennas, $K=8$ \acp{UE}, and $N=16$ BD-RIS elements.}
\label{fig:FRIS}
\end{figure*}

In Fig.~\ref{fig:FRIS} we compare the achievable sum capacity ratio, which is the counterpart of the achievable capacity ratio for the multi-user case, when feedbacking the \ac{FRIS} unitary capacity achieving configuration from \cite{bartoli}. We have averaged the results throughout $10^4$ realizations of the cascaded channels $\bs{H}_1$ and $\bs{H}_2$, both modeled as standard \ac{IID} Rayleigh fading channels. The \ac{SNR} is fixed to $10$ dB, the number of transmit antennas to $M_\mrm{T}=8$, and the number of receive antennas to $M_\mrm{R}=16$, while we considered $N\in\{8,16\}$ \ac{FRIS} elements, which determine the size of the unitary matrices to be sent. As we can see from Fig.~\ref{fig:AWGN_FRIS}, the proposed \ac{DEP} outperforms the other approaches by a great margin under the \ac{AWGN} feedback channel, similar to what happened in the previous \ac{CSI} feedback scenario. In Fig.~\ref{fig:q_FRIS}, we also see the same trend previously observed for the uniformly quantized feedback, where in this case the proposed \ac{DEP} with overrange shows the highest potential since RIS-related scenarios are characterized by a large number of passive reconfigurable elements, which favor the proposed approach. Note that the overrange has been increased manually to the point before the performance starts to degrade, but in future work we may gain more insights by studying the distribution of the entries of the proposed \ac{DEP}.

\subsection{Decentralized hybrid beamforming sum capacity ratio}
We end by analyzing the performance of the considered approaches in the generalized decentralized hybrid beamforming scenario described in Section~\ref{ssec:dec}, based on \cite{asilomar24}. We assume that the scheme from \cite{asilomar24} for computing the decentralized analog filters is implemented at the \ac{BBU}, and the resulting unitary matrices are shared with the decentralized modules using the previous approaches. In this case, the achievable sum capacity ratio is computed with respect to the channel sum capacity, as in \cite{asilomar24}, and averaged over $10^3$ realizations of an \ac{IID} Rayleigh fading channel. This means that, even if the unitary matrices are perfectly reconstructed, the scheme may not converge to the full capacity unless the parameter combination allows for information-lossless processing \cite{wax_journal}.

\begin{figure*}[h]
\begin{subfigure}{\columnwidth}
  \centering
    \includegraphics[scale=0.55]{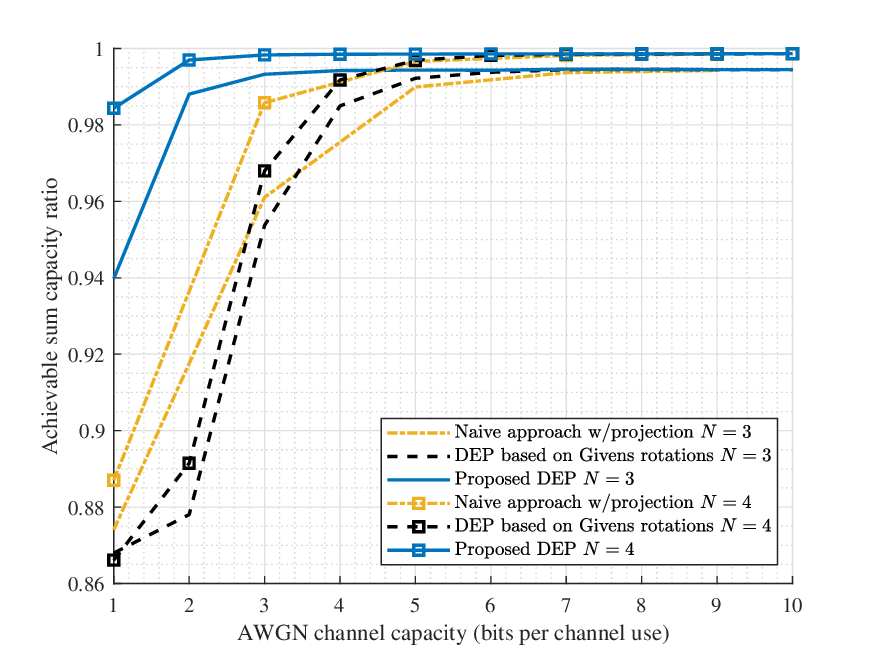}
  \caption{\ac{AWGN} sharing channel.}
  \label{fig:AWGN_WAX}
  \end{subfigure}
\begin{subfigure}{\columnwidth}
  \centering
    \includegraphics[scale=0.55]{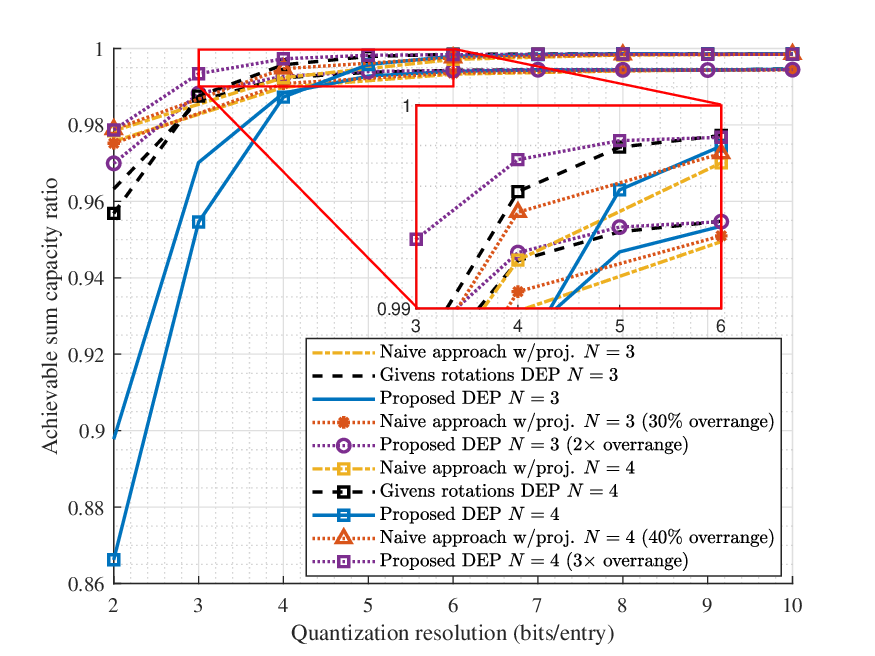}
  \caption{Quantized sharing.}
  \label{fig:q_WAX}
\end{subfigure}
  \caption{Achievable sum capacity ratio for decentralized analog beamformer sharing in \cite{asilomar24} with $M=12$ \ac{BS} antennas, $K=6$ \acp{UE}, and $T=8$ inputs to the \ac{BBU}.}
  \label{fig:WAX}
\end{figure*}

In Fig.~\ref{fig:AWGN_WAX} we see again that the proposed \ac{DEP} is the most efficient way to transmit unitary matrices under the \ac{AWGN} channel, allowing to halve the required data rate with respect to the other approaches. In the case of quantized data from Fig.~\ref{fig:q_WAX}, we see that, even for this small matrix sizes characterizing the size of the decentralized filters, the proposed \ac{DEP} accepts a reasonable amount of overrange, which allows it to outperform the Givens rotations \ac{DEP}.

\section{Conclusions}\label{section:conc}
In this work, we have addressed the transmission and storage of unitary matrices by mapping these matrices to a sequence of bounded real numbers with minimum dimension. To this end, we have derived a \ac{DEP}, and we have characterized an interval bounding the real numbers associated to this \ac{DEP}. We have presented several applications from the field of wireless communications that could benefit from employing this technique. We have also shown how to adjust the \ac{DEP} when considering extra matrix constraints that may appear in some of its applications, showcasing the adaptability and practicality of our approach. Finally, we have performed a numerical analysis under \ac{AWGN} channels, as well as under uniform quantization, demonstrating the potential of the proposed approach in general applications, including the mentioned ones from wireless communications. In the case of uniform quantization, the proposed approach requires adjusting the quantization range allowing some overrange for the input values with respect to the derived bounds. In this sense, an interesting direction for future work is to study alternative quantization approaches for the proposed \ac{DEP} so as to better exploit its potential gains.
\bibliographystyle{IEEEtran}
\bibliography{IEEEabrv,bibliography}

%

\end{document}